\documentclass{ws-ijgmmp}
\usepackage{amssymb}

\begin{document}


%
\catchline{}{}{}{}{}
%

\title{Quantum Geometry and Wild embeddings as quantum states}

\author{Torsten Asselmeyer-Maluga}

\address{German Aerospacecenter, Rutherfordstr. 2, D-12489 Berlin, Germany\\ 
\email{torsten.asselmeyer-maluga@dlr.de}}

\author{Jerzy Kr{\'o}l}

\address{University of Silesia, Institute of Physics, ul. Uniwesytecka 4, 40-007 Katowice, Poland\\
\email{iriking@wp.pl}}

\maketitle

\begin{history}
\received{(Day Month Year)}
\revised{(Day Month Year)}
\end{history}

\begin{abstract}
In this paper we discuss wild embeddings like Alexanders horned ball
and relate them to fractal spaces. We build a $C^{\star}$-algebra
corresponding to a wild embedding. We argue that a wild embedding
is the result of a quantization process applied to a tame embedding.
Therefore quantum states are directly the wild embeddings. Then we
give an example of a wild embedding in the 4-dimensional spacetime.
We discuss the consequences for cosmology.
\end{abstract}


\tableofcontents{}

\section{Introduction}

General relativity (GR) has changed our understanding of spacetime.
In parallel, the appearance of quantum field theory (QFT) has modified
our view of particles, fields and the measurement process. The usual
approach for the unification of QFT and GR, to a quantum gravity,
starts with a proposal to quantize GR and its underlying structure,
spacetime. There is an unique opinion in the community about the relation
between geometry and quantum theory: The geometry as used in GR is
classical and should emerge from a quantum gravity in the limit (Planck's
constant tends to zero). Most theories went a step further and try
to get a spacetime from quantum theory. But what happens if this prerequisite
is wrong? Is it possible to derive the quantization procedure from
the structure of space and time? What is the geometrical representation
of a quantum state? This paper try to answer partly these questions.

All approaches of quantum gravity have problems with the construction
of the state space. If we assume that the spacetime has the right
properties for a spacetime picture of quantum gravity then the quantum
state must be part of the spacetime or must be geometrically realized
in the spacetime. Consider (as in geometrodynamics) a 3-sphere $S^{3}$
with metric $g$. This metric (as state of GR) is modeled on $S^{3}$
at every 3-dimensional subspace. If $g$ is a metric of a homogeneous
space then one can choose a small coordinate patch. But if $g$ is
inhomogeneous then one can use a diffeomorphism to ''concentrate''
the inhomogeneity at a chart. Now one combines these infinite charts
(we consider only metrics up to diffeomorphisms) into a 3-sphere but
without destroying the infinite charts by a diffeomorphism. Then we
obtain a model of a quantum state. But as we argue in this paper,
wild embeddings are the right structure to realize this idea. A wild
embedding cannot be undone by a diffeomorphism of the embedding space.
But it is non-trivial and (in most cases) determined by its complement.
If we assume that wild embeddings are the quantum states then we must
obtain the wild embedding by a quantization process. In this paper
we will construct a $C^{*}-$algebra and its enveloping von Neumann
algebra in the next section. Then we will show that closed curves
(used to construct the $C^{*}-$algebra) are the observables forming
a Poisson algebra for a tame embedding. The deformation quantization
(so-called Drinfeld-Turaev-quantization) of this Poisson algebra leads
to skein spaces used in knot theory. For the example of Alexanders
horned ball we are able to show that the corresponding skein space
and the enveloping von Neumann algebra are the same. Therefore the
wild embedding (Alexanders horned ball) can be seen as a quantization
of a tame embedding, i.e. it is a quantum state. Wild embeddings occur
generically in non-compact 4-manifolds with an exotic smoothness structure
like $S^{3}\times\mathbb{R}$. In the last section we discuss also
the cosmological consequences of this approach.

\section{From wild embeddings to fractal spaces}

In this section we define wild and tame embeddings and construct a
$C^{*}-$algebra associated to a wild embedding. The example of Alexanders
horned ball is discussed.

\subsection{Wild and tame embeddings\label{sub:Wild-and-tame-embed}}

We call a map $f:N\to M$ between two topological manifolds an embedding
if $N$ and $f(N)\subset M$ are homeomorphic to each other. From
the differential-topological point of view, an embedding is a map
$f:N\to M$ with injective differential on each point (an immersion)
and $N$ is diffeomorphic to $f(N)\subset M$. An embedding $i:N\hookrightarrow M$
is \emph{tame} if $i(N)$ is represented by a finite polyhedron homeomorphic
to $N$. Otherwise we call the embedding \emph{wild}. There are famous
wild embeddings like Alexanders horned sphere or Antoine's necklace.
In physics one uses mostly tame embeddings but as Cannon mentioned
in his overview \cite{Can:78}, one needs wild embeddings to understand
the tame one. As shown by us \cite{AsselmeyerKrol2009}, wild embeddings
are needed to understand exotic smoothness. As explained in \cite{Can:78}
by Cannon, tameness is strongly connected to another topic: decomposition
theory (see the book \cite{Daverman1986}). 

Two embeddings $f,g:N\to M$ are said to be isotopic, if there exists
a homeomorphism $F:M\times[0,1]\to M\times[0,1]$ such that 
\begin{enumerate}
\item $F(y,0)=(y,0)$ for each $y\in M$ (i.e. $F(.,0)=id_{M}$)
\item $F(f(x),1)=g(x)$ for each $x\in N$, and
\item $F(M\times\left\{ t\right\} )=M\times\left\{ t\right\} $ for each
$t\in[0,1]$.
\end{enumerate}
If only the first two conditions can be fulfilled then one call it
concordance. Embeddings are usually classified by isotopy. An important
example is the embedding $S^{1}\to\mathbb{R}^{3}$, known as knot,
where different knots are different isotopy classes of the embedding
$S^{1}\to\mathbb{R}^{3}$.

\subsection{Wild embeddings and perfect groups}

Wild embeddings are important to understand usual embeddings. Consider
a closed curve in the plane. This curve divides the plane into an
interior and an exterior area (by the Jordan curve theorem). But what
about one dimension higher, i.e. consider the embedding $S^{2}\to\mathbb{R}^{3}$?
Alexander was the first who constructed a counterexample to a generalized
Jordan curve theorem, Alexanders horned sphere \cite{Alex:24}, as
wild embedding $I:D^{3}\to\mathbb{R}^{3}$. The main property of this
wild object $D_{W}^{3}=I(D^{3})$ is the non-simple connected complement
$\mathbb{R}^{3}\setminus D_{W}^{3}$. This property is a crucial point
of the following discussion. Given an embedding $I:D^{3}\to\mathbb{R}^{3}$
which induces a decomposition $\mathbb{R}^{3}=I(D^{3})\cup(\mathbb{R}^{3}\setminus I(D^{3}))$.
In case, the embedding is tame, the image $I(D^{3})$ is given by
a finite complex and every part of the decomposition is contractable,
i.e. especially $\pi_{1}(\mathbb{R}^{3}\setminus I(D^{3}))=0$. For
a wild embedding, $I(D^{3})$ is an infinite complex (but contractable).
The complement $\mathbb{R}^{3}\setminus I(D^{3})$ is given by a sequence
of spaces so that $\mathbb{R}^{3}\setminus I(D^{3})$ is non-simple
connected (otherwise the embedding must be tame). If $\mathbb{R}^{3}\setminus I(D^{3})$
has the homology of a point (that is true for every embedding) then
$\pi_{1}(\mathbb{R}^{3}\setminus I(D^{3}))$ is non-trivial whereas
its abelization $H_{1}(\mathbb{R}^{3}\setminus I(D^{3}))=0$ vanishes.
Therefore $\pi_{1}$ is generated by the commutator subgroup $[\pi_{1},\pi_{1}]$
with $[a,b]=aba^{-1}b^{-1}$ for two elements $a,b\in\pi_{1}$, i.e.
$\pi_{1}$ is a perfect group.

As a warm up we will consider wild embeddings of spheres $S^{n}$
into spheres $S^{m}$ equivalent to embeddings of $\mathbb{R}^{n}$
into $\mathbb{R}^{m}$relative to the infinity $\infty$ point or
to relative embeddings of $D^{n}$ into $D^{m}$ (relative to its
boundary). From the physical point of view, we have the embedding
of branes (seen as topological objects of a trivial type like $\mathbb{R}^{n},S^{n}$
or $D^{n}$) into the Euclidean space $\mathbb{R}^{m}$. Lets start
with the case of a finite $k-$dimensional polyhedron $K^{k}$ (i.e.
a piecewise-linear version of a $k-$disk $D^{k}$). Consider the
wild embedding $i:K\to S^{n}$ with $0\leq k\leq n-3$ and $n\geq7$.
Then, as shown in \cite{FerryPedersenVogel1989}, the complement $S^{n}\setminus i(K)$
is non-simple connected with a countable generated (but not finitely
presented) fundamental group $\pi_{1}(S^{n}\setminus i(K))=\pi$.
Furthermore, the group $\pi$ is perfect (i.e. generated by the commutator
subgroup $[\pi,\pi]=\pi$ implying $H_{1}(\pi)=0$) and $H_{2}(\pi)=0$
($\pi$ is called a superperfect group). Again, $\pi$ is a group
where every element $x\in\pi$ can be generated by a commutator $x=[a,b]=aba^{-1}b^{-1}$
(including the trivial case $x=a,\: b=e$). By using geometric group
theory, one can represent $\pi$ by a grope (or generalized disk,
see Cannon \cite{Can:79}), i.e. a hierarchical object with the same
fundamental group as $\pi$. This group is not finite in case of a
wild embedding.

\subsection{$C^{*}-$algebras associated to wild embeddings and idempotents\label{sub:C*-algebra-wild-embedding_idempotent}}

Let $I:K^{n}\to\mathbb{R}^{n+k}$ be a wild embedding of codimension
$k$ with $k=0,1,2$. In the following we assume that the complement
$\mathbb{R}^{n+k}\setminus I(K^{n})$ is non-trivial, i.e. $\pi_{1}(\mathbb{R}^{n+k}\setminus I(K^{n}))=\pi\not=1$.
Now we define the $C^{*}-$algebra $C^{*}(\mathcal{G},\pi$) associated
to the complement $\mathcal{G}=\mathbb{R}^{n+k}\setminus I(K^{n})$
with group $\pi=\pi_{1}(\mathcal{G})$. If $\pi$ is non-trivial then
this group is not finitely generated. The construction of wild embeddings
is usually given by an infinite construction%
\footnote{This infinite construction is necessary to obtain an infinite polyhedron,
the defining property of a wild embedding.%
} (see Antoine\textquoteright{}s necklace or Alexanders horned sphere).
From an abstract point of view, we have a decomposition of $\mathcal{G}$
by an infinite union
\[
\mathcal{G}=\bigcup_{i=0}^{\infty}C_{i}
\]
of 'level sets' $C_{i}$. Then every element $\gamma\in\pi$ lies
(up to homotopy) in a finite union of levels. 

The basic elements of the $C^{*}-$algebra $C^{*}(\mathcal{G},\pi$)
are smooth half-densities with compact supports on $\mathcal{G}$,
$f\in C_{c}^{\infty}(\mathcal{G},\Omega^{1/2})$, where $\Omega_{\gamma}^{1/2}$
for $\gamma\in\pi$ is the one-dimensional complex vector space of
maps from the exterior power $\Lambda^{k}L$ ($\dim L=k$), of the
union of levels $L$ representing $\gamma$, to $\mathbb{C}$ such
that 
\[
\rho(\lambda\nu)=|\lambda|^{1/2}\rho(\nu)\qquad\forall\nu\in\Lambda^{2}L,\lambda\in\mathbb{R}\:.
\]
For $f,g\in C_{c}^{\infty}(\mathcal{G},\Omega^{1/2})$, the convolution
product $f*g$ is given by the equality
\[
(f*g)(\gamma)=\intop_{\gamma_{1}\circ\gamma_{2}=\gamma}f(\gamma_{1})g(\gamma_{2})
\]
with the group operation $\gamma_{1}\circ\gamma_{2}$ in $\pi$. Then
we define via $f^{*}(\gamma)=\overline{f(\gamma^{-1})}$ a $*$operation
making $C_{c}^{\infty}(\mathcal{G},\Omega^{1/2})$ into a $*$algebra.
Each level set $C_{i}$ consists of simple pieces (in case of Alexanders
horned sphere, we will explain it below) denoted by $T$. For these
pieces, one has a natural representation of $C_{c}^{\infty}(\mathcal{G},\Omega^{1/2})$
on the $L^{2}$ space over $T$. Then one defines the representation
\[
(\pi_{x}(f)\xi)(\gamma)=\intop_{\gamma_{1}\circ\gamma_{2}=\gamma}f(\gamma_{1})\xi(\gamma_{2})\qquad\forall\xi\in L^{2}(T),\forall x\in\gamma.
\]
The completion of $C_{c}^{\infty}(\mathcal{G},\Omega^{1/2})$ with
respect to the norm 
\[
||f||=\sup_{x\in\mathcal{G}}||\pi_{x}(f)||
\]
makes it into a $C^{*}$algebra $C_{c}^{\infty}(\mathcal{G},\pi$).
Finally we are able to define the $C^{*}-$algebra associated to the
wild embedding: \begin{definition} Let $j:K\to S^{n}$ be a wild
embedding with $\pi=\pi_{1}(S^{n}\setminus j(K))$ as fundamental
group of the complement $M(K,j)=S^{n}\setminus j(K)$. The $C^{*}-$algebra
$C_{c}^{\infty}(K,j)$ associated to the wild embedding is defined
to be $C_{c}^{\infty}(K,j)=C_{c}^{\infty}(\mathcal{G},\pi)$ the $C^{*}-$algebra
of the complement $\mathcal{G}=S^{n}\setminus j(K)$ with group $\pi$.
\end{definition} Among all elements of the $C^{*}$ algebra, there
are distinguished elements, idempotent operators or projectors having
a geometric interpretation. For later use, we will construct them
explicitly (we follow \cite{Connes94} sec. $II.8.\beta$ closely).
Let $j(K)\subset S^{n}$ be the wild submanifold. A small tubular
neighborhood $N'$ of $j(K)$ in $S^{n}$ defines the corresponding
$C^{*}$algebra $C_{c}^{\infty}(N',\pi_{1}(S^{n}\setminus N'))$ is
isomorphic to $C_{c}^{\infty}(\mathcal{G},\pi_{1}(S^{n}\setminus j(K)))\otimes\mathcal{K}$
with $\mathcal{K}$ the $C^{*}$ algebra of compact operators. In
particular it contains an idempotent $e=e^{2}=e^{*}$, $e=1_{N}\otimes f\in C_{c}^{\infty}(\mathcal{G},\pi_{1}(S^{n}\setminus j(K)))\otimes\mathcal{K}$
, where $f$ is a minimal projection in $\mathcal{K}$. It induces
an idempotent in $C_{c}^{\infty}(\mathcal{G},\pi_{1}(S^{n}\setminus j(K)))$.
By definition, this idempotent is given by a closed curve in the complement
$S^{n}\setminus j(K)$.

\subsection{Example: Alexanders horned ball as fractal space\label{sub:Example-Alexanders-horned-ball}}

\begin{figure}
\includegraphics[scale=0.3]{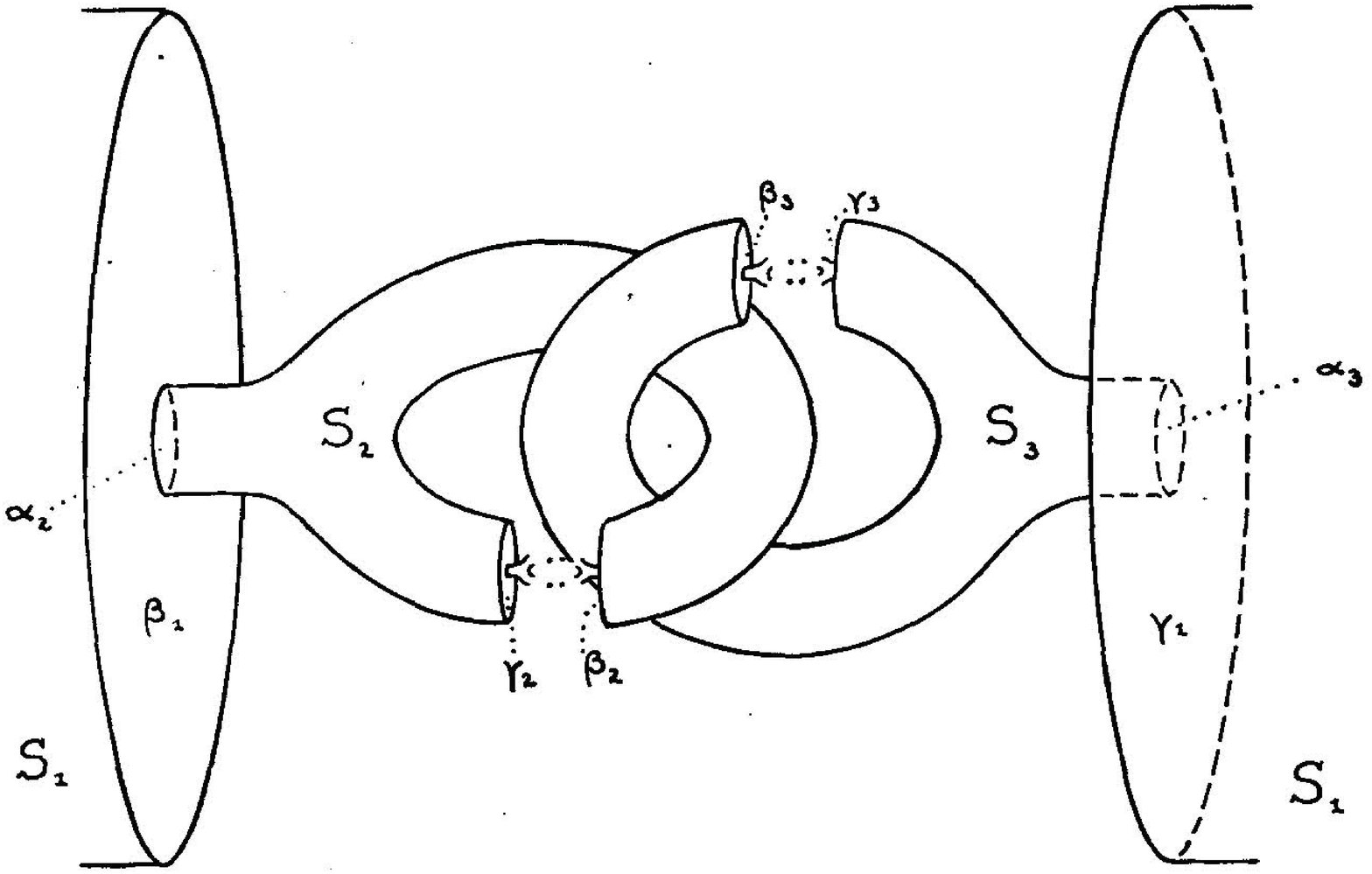}

\caption{linking the two manifolds (denoted by $S_{2}$ and $S_{3}$)
\label{fig:linking-two-Y}}

\end{figure}

In this subsection we will construct Alexanders horned ball (originally
described in \cite{Alex:24}) as an example of a wild embedding $D^{3}\to S^{3}$.
The construction needs an infinite number of levels $C_{i}$. The
$0$th level $C_{0}$ is a solid cylinder (or annulus) $D^{2}\times[0,1]$
with the two ends $D^{2}\times\left\{ 0\right\} $ and $D^{2}\times\left\{ 1\right\} $.
For the first level $C_{1}$, consider a pair of $Y=\left(D^{2}\times[0,1]\right)\setminus D^{2}$
(the thicken letter 'Y') with three ends $D^{2}\times\left\{ 0\right\} $,
$D^{2}\times\left\{ \frac{1}{2}\right\} $ and $D^{2}\times\left\{ 1\right\} $.
Now glue $D^{2}\times\left\{ \frac{1}{2}\right\} $ of one $Y$ to
one end of $C_{0}$ and do the same for the other $Y$. Then link
the two $Y$ (as shown in fig. \ref{fig:linking-two-Y} which is repainted
from \cite{Alex:24}) and repeat
the procedure for every pair of ends $D^{2}\times\left\{ 0\right\} $
and $D^{2}\times\left\{ 1\right\} $ of $Y$. 
 As explained above, the infinite union 
\[
C=\bigcup_{i=0}^{\infty}C_{i}
\]
 of all levels is the image $I(D^{3})$ of the wild embedding $I:D^{3}\to S^{3}$,
known as Alexanders horned ball $A=I(D^{3})$. 

Wild embeddings are known to be fractals \cite{WildFractal1992}.
It is also possible to give an estimate for the fractal dimension
for the boundary of Alexander's horned ball, i.e. for a wildly embedded
2-sphere. First assume that the next level is twice smaller then the
previous one (scale $1:2=1:m$). Secondly, in every of the three direction,
one has two copies of the whole object, i.e. one obtains $n=6$ disjoint
objects. Then one obtains for the Hausdorff dimension
\[
D=\frac{\ln n}{\ln m}=\frac{\ln6}{\ln2}\approx2.584925...
\]
a greater value then the usual dimension of the 2-sphere. We remark
that this simple calculation is only a rough estimate to demonstrate
the fact that wild embeddings are fractals.

\begin{figure}
\includegraphics{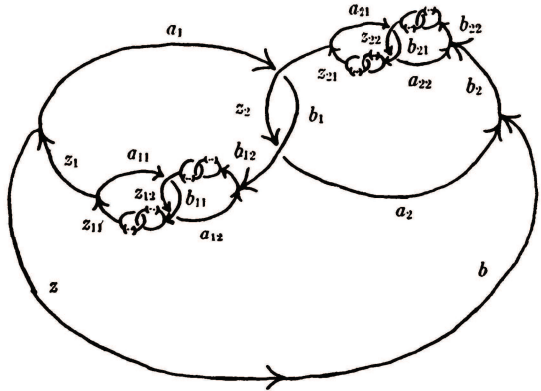}

\caption{schematic picture for the first three levels 
\label{fig:schematical-picture-for-A}}

\end{figure}

For an impression of Alexanders horned ball $A$, we have to consider
the complement $S^{3}\setminus A$. Main instrument in this context
is the fundamental group $\pi_{1}(S^{3}\setminus A)$ of the complement.
It was constructed in \cite{FoxBlankinship1950} by using a schematic
picture of $A$ (the spine in modern notation), see Fig. \ref{fig:schematical-picture-for-A}(repainted from \cite{FoxBlankinship1950}).
 Let $\alpha$ be a finite sequence $\alpha=\alpha_{1}\alpha_{2}\cdots\alpha_{l}$
of length $l(\alpha)=l$ over the alphabet $\alpha_{i}=1,2$. Then
the fundamental group is given by
\[
\pi_{1}(S^{3}\setminus A))=\left\{ z_{\alpha},l(\alpha)\geq0\,|\: z_{\alpha}=[z_{\alpha1},z_{\alpha2}^{-1}]\right\} 
\]
with the generators $z_{\alpha}$ by using the group commutator $[a,b]=aba^{-1}b^{-1}$.
The group $\pi_{1}(S^{3}\setminus A)$ is a locally free group of
infinite rank which is perfect. But the last property implies that
this group has the infinite conjugacy class property (icc), i.e. only
the identity element has a finite conjugacy class. This property has
a tremendous impact on the $C^{*}-$algebra \cite{Connes94} and its
enveloping von Neumann algebra:\\
\emph{The enveloping von Neumann algebra $W(C,\pi_{1}(S^{3}\setminus A))$
of the $C^{*}-$algebra 
\[
C_{c}^{\infty}(C,\pi_{1}(S^{3}\setminus A))
\]
 for the wild embedding $A$ is the hyperfinite factor $I\! I_{1}$
algebra.}

\section{Quantum states from wild embeddings}

In this section we will describe a way from a (classical) Poisson
algebra to a quantum algebra by using deformation quantization. Therefore
we will obtain a positive answer to the question: Does the $C^{*}-$algebra
of a wild (specific) embedding comes from a (deformation) quantization?
Of course, this question cannot be answered in most generality, i.e.
we use the example of Alexanders horned ball. But for this example
we will show that the enveloping von Neumann algebra of this wild
embedding (Alexanders horned ball) is the result of a deformation
quantization using the classical Poisson algebra (of closed curves)
of the tame embedding. This result shows two things: the wild embedding
can be seen as a quantum state and the classical state is a tame embedding.
We conjecture that this result can be generalized to all other wild
embeddings.

\subsection{Embeddings and observables}

Let $i:K\to S^{n}$ be an embedding of a polyhedron $K$. If we have
a fixed geometry of $S^{n}$ then we obtain also an induced geometry
of $i(K)$ by using the embedding. Without loss of generality, we
can assume a connection $\Gamma_{S^{n}}$ of constant curvature on
$S^{n}$. By a pullback $\Gamma_{K}=i^{*}\Gamma_{S^{n}}$ we obtain
a connection of $i(K)\subset S^{n}$. Using the Ambrose-Singer theorem
we obtain the curvature via the holonomy
\[
\ointop_{\gamma}\,\Gamma_{K}
\]
along a closed curve $\gamma\in i(K)$. In case of a constant curvature,
this integral depends only on the homotopy class of the curve $\gamma$.
But also for every piece having a constant curvature, we can choose
a closed curve (in this piece) up to homotopy. Therefore, the closed
curve is an important ingredient to obtain the curvature. 

Now we consider the complement $S^{n}\setminus i(K)$ together with
the fundamental group $\pi=\pi_{1}\left(S^{n}\setminus i(K)\right)$.
An element of $\pi$ is a closed curve surrounding $i(K)$. The curve
can be projected to $i(K)$. In most cases, the group $\pi$ is the
trivial group and we obtain the algebra $C_{c}^{\infty}(S^{n}\setminus i(K),\pi)=\mathbb{C}$,
the center of the enveloping von Neumann algebra. The generator of
$\pi$ is a contractable closed curve used to determine the curvature
(and therefore the geometry). So, if we have a fixed geometry of the
embedding space (preferable constant curvature) then we can construct
the curvature (up to diffeomorphisms) by choosing a closed curve.
In this case, \emph{the closed is an observable}.%
\footnote{This approach is not completely new. The spin network of Loop quantum
gravity is also an expression to obtain the geometry using the holonomy
of a connection along the network.%
} This concept agrees with the construction of an idempotent in the
$C^{*}-$algebra (see subsection \ref{sub:C*-algebra-wild-embedding_idempotent})
seen as closed curve surrounding the complement of the wild embedding.

In the example of Alexanders horned ball, we also obtain a wild embedding
of a 2-sphere (as boundary of the ball) into the 3-space. This case,
the wild embedding of a surface into the 3-space, will be the generic
case for the following. We will motivate the choice in section \ref{sec:Wild-embeddings-exotic-4MF}.

\subsection{Intermezzo 1: The observable algebra and its Poisson structure\label{sub:The-observable-algebra}}

In this section we will describe the formal structure of a classical
theory coming from the algebra of observables using the concept of
a Poisson algebra. In quantum theory, an observable is represented
by a hermitean operator having the spectral decomposition via projectors
or idempotent operators. The coefficient of the projector is the eigenvalue
of the observable or one possible result of a measurement. At least
one of these projectors represent (via the GNS representation) a quasi-classical
state. Thus to construct the substitute of a classical observable
algebra with Poisson algebra structure we have to concentrate on the
idempotents in the $C^{*}$ algebra. Now we will see that the set
of closed curves on a surface has the structure of a Poisson algebra.
Let us start with the definition of a Poisson algebra. \begin{definition}

Let $P$ be a commutative algebra with unit over $\mathbb{R}$ or
$\mathbb{C}$. A \emph{Poisson bracket} on $P$ is a bilinearform
$\left\{ \:,\:\right\} :P\otimes P\to P$ fulfilling the following
3 conditions:

anti-symmetry $\left\{ a,b\right\} =-\left\{ b,a\right\} $

Jacobi identity $\left\{ a,\left\{ b,c\right\} \right\} +\left\{ c,\left\{ a,b\right\} \right\} +\left\{ b,\left\{ c,a\right\} \right\} =0$

derivation $\left\{ ab,c\right\} =a\left\{ b,c\right\} +b\left\{ a,c\right\} $.\\
Then a \emph{Poisson algebra} is the algebra $(P,\{\,,\,\})$.\end{definition}

Now we consider a surface $S$ together with a closed curve $\gamma$.
Additionally we have a Lie group $G$ given by the isometry group.
The closed curve is one element of the fundamental group $\pi_{1}(S)$.
From the theory of surfaces we know that $\pi_{1}(S)$ is a free abelian
group. Denote by $Z$ the free $\mathbb{K}$-module ($\mathbb{K}$
a ring with unit) with the basis $\pi_{1}(S)$, i.e. $Z$ is a freely
generated $\mathbb{K}$-module. Recall Goldman's definition of the
Lie bracket in $Z$ (see \cite{Goldman1984}). For a loop $\gamma:S^{1}\to S$
we denote its class in $\pi_{1}(S)$ by $\left\langle \gamma\right\rangle $.
Let $\alpha,\beta$ be two loops on $S$ lying in general position.
Denote the (finite) set $\alpha(S^{1})\cap\beta(S^{1})$ by $\alpha\#\beta$.
For $q\in\alpha\#\beta$ denote by $\epsilon(q;\alpha,\beta)=\pm1$
the intersection index of $\alpha$ and $\beta$ in $q$. Denote by
$\alpha_{q}\beta_{q}$ the product of the loops $\alpha,\beta$ based
in $q$. Up to homotopy the loop $(\alpha_{q}\beta_{q})(S^{1})$ is
obtained from $\alpha(S^{1})\cup\beta(S^{1})$ by the orientation
preserving smoothing of the crossing in the point $q$. Set 
\begin{equation}
[\left\langle \alpha\right\rangle ,\left\langle \beta\right\rangle ]=\sum_{q\in\alpha\#\beta}\epsilon(q;\alpha,\beta)(\alpha_{q}\beta_{q})\quad.\label{eq:Lie-bracket-loops}
\end{equation}
According to Goldman \cite{Goldman1984}, Theorem 5.2, the bilinear
pairing $[\,,\,]:Z\times Z\to Z$ given by (\ref{eq:Lie-bracket-loops})
on the generators is well defined and makes $Z$ to a Lie algebra.
The algebra $Sym(Z)$ of symmetric tensors is then a Poisson algebra
(see Turaev \cite{Turaev1991}).

The whole approach seems natural for the construction of the Lie algebra
$Z$ but the introduction of the Poisson structure is an artificial
act. From the physical point of view, the Poisson structure is not
the essential part of classical mechanics. More important is the algebra
of observables, i.e. functions over the configuration space forming
the Poisson algebra. Thus we will look for the algebra of observables
in our case. For that purpose, we will look at geometries over the
surface. By the uniformization theorem of surfaces, there is three
types of geometrical models: spherical $S^{2}$, Euclidean $\mathbb{E}^{2}$
and hyperbolic $\mathbb{H}^{2}$. Let $\mathcal{M}$ be one of these
models having the isometry group $Isom(\mathcal{M})$. Consider a
subgroup $H\subset Isom(\mathcal{M})$ of the isometry group acting
freely on the model $\mathcal{M}$ forming the factor space $\mathcal{M}/H$.
Then one obtains the usual (closed) surfaces $S^{2}$, $\mathbb{R}P^{2}$,
$T^{2}$ and its connected sums like the surface of genus $g$ ($g>1$).
For the following construction we need a group $G$ containing the
isometry groups of the three models. Furthermore the surface $S$
is part of a 3-manifold and for later use we have to demand that $G$
has to be also a isometry group of 3-manifolds. According to Thurston
\cite{Thu:97} there are 8 geometric models in dimension 3 and the
largest isometry group is the hyperbolic group $PSL(2,\mathbb{C})$
isomorphic to the Lorentz group $SO(3,1).$ It is known that every
representation of $PSL(2,\mathbb{C})$ can be lifted to the spin group
$SL(2,\mathbb{C})$. Thus the group $G$ fulfilling all conditions
is identified with $SL(2,\mathbb{C})$. This choice fits very well
with the 4-dimensional picture.

Now we introduce a principal $G$ bundle on $S$, representing a geometry
on the surface. This bundle is induced from a $G$ bundle over $S\times[0,1]$
having always a flat connection. Alternatively one can consider a
homomorphism $\pi_{1}(S)\to G$ represented as holonomy functional
\begin{equation}
hol(\omega,\gamma)=\mathcal{P}\exp\left(\int\limits _{\gamma}\omega\right)\in G\label{eq:holonomy-definition}
\end{equation}
with the path ordering operator $\mathcal{P}$ and $\omega$ as flat
connection (i.e. inducing a flat curvature $\Omega=d\omega+\omega\wedge\omega=0$).
This functional is unique up to conjugation induced by a gauge transformation
of the connection. Thus we have to consider the conjugation classes
of maps
\[
hol:\pi_{1}(S)\to G
\]
forming the space $X(S,G)$ of gauge-invariant flat connections of
principal $G$ bundles over $S$. Now (see \cite{Skovborg2006}) we
can start with the construction of the Poisson structure on $X(S,G).$
The construction based on the Cartan form as the unique bilinearform
of a Lie algebra. As discussed above we will use the Lie group $G=SL(2,\mathbb{C})$
but the whole procedure works for every other group too. Now we consider
the standard basis
\[
X=\left(\begin{array}{cc}
0 & 1\\
0 & 0
\end{array}\right)\quad,\qquad H=\left(\begin{array}{cc}
1 & 0\\
0 & -1
\end{array}\right)\quad,\qquad Y=\left(\begin{array}{cc}
0 & 0\\
1 & 0
\end{array}\right)
\]
of the Lie algebra $sl(2,\mathbb{C})$ with $[X,Y]=H,\,[H,X]=2X,\,[H,Y]=-2Y$.
Furthermore there is the bilinearform $B:sl_{2}\otimes sl_{2}\to\mathbb{C}$
written in the standard basis as 
\[
\left(\begin{array}{ccc}
0 & 0 & -1\\
0 & -2 & 0\\
-1 & 0 & 0
\end{array}\right)
\]
Now we consider the holomorphic function $f:SL(2,\mathbb{C})\to\mathbb{C}$
and define the gradient $\delta_{f}(A)$ along $f$ at the point $A$
as $\delta_{f}(A)=Z$ with $B(Z,W)=df_{A}(W)$ and 
\[
df_{A}(W)=\left.\frac{d}{dt}f(A\cdot\exp(tW))\right|_{t=0}\quad.
\]
The calculation of the gradient $\delta_{tr}$ for the trace $tr$
along a matrix 
\[
A=\left(\begin{array}{cc}
a_{11} & a_{12}\\
a_{21} & a_{22}
\end{array}\right)
\]
 is given by
\[
\delta_{tr}(A)=-a_{21}Y-a_{12}X-\frac{1}{2}(a_{11}-a_{22})H\quad.
\]
Given a representation $\rho\in X(S,SL(2,\mathbb{C}))$ of the fundamental
group and an invariant function $f:SL(2,\mathbb{C})\to\mathbb{R}$
extendable to $X(S,SL(2,\mathbb{C}))$. Then we consider two conjugacy
classes $\gamma,\eta\in\pi_{1}(S)$ represented by two transversal
intersecting loops $P,Q$ and define the function $f_{\gamma}:X(S,SL(2,\mathbb{C})\to\mathbb{C}$
by $f_{\gamma}(\rho)=f(\rho(\gamma))$. Let $x\in P\cap Q$ be the
intersection point of the loops $P,Q$ and $c_{x}$ a path between
the point $x$ and the fixed base point in $\pi_{1}(S)$. Then we
define $\gamma_{x}=c_{x}\gamma c_{x}^{-1}$ and $\eta_{x}=c_{x}\eta c_{x}^{-1}$.
Finally we get the Poisson bracket
\[
\left\{ f_{\gamma},f'_{\eta}\right\} =\sum_{x\in P\cap Q}sign(x)\: B(\delta_{f}(\rho(\gamma_{x})),\delta_{f'}(\rho(\eta_{x})))\quad,
\]
where $sign(x)$ is the sign of the intersection point $x$. Thus,

The space $X(S,SL(2,\mathbb{C}))$ has a natural Poisson structure
(induced by the bilinear form (\ref{eq:Lie-bracket-loops}) on the
group) and the Poisson algebra \emph{$(X(S,SL(2,\mathbb{C}),\left\{ \,,\,\right\} )$}
of complex functions over them is the algebra of observables.

\subsection{Intermezzo 2: Drinfeld-Turaev Quantization\label{sub:Drinfeld-Turaev-Quantization}}

Now we introduce the ring $\mathbb{C}[[h]]$ of formal polynomials
in $h$ with values in $\mathbb{C}$. This ring has a topological
structure, i.e. for a given power series $a\in\mathbb{C}[[h]]$ the
set $a+h^{n}\mathbb{C}[[h]]$ forms a neighborhood. Now we define
\begin{definition}

A \emph{Quantization} of a Poisson algebra $P$ is a $\mathbb{C}[[h]]$
algebra $P_{h}$ together with the $\mathbb{C}$-algebra isomorphism
$\Theta:P_{h}/hP\to P$ so that

1. the module $P_{h}$ is isomorphic to $V[[h]]$ for a $\mathbb{C}$
vector space $V$

2. let $a,b\in P$ and $a',b'\in P_{h}$ be $\Theta(a)=a'$, $\Theta(b)=b'$
then
\[
\Theta\left(\frac{a'b'-b'a'}{h}\right)=\left\{ a,b\right\} 
\]
\end{definition}

One speaks of a deformation of the Poisson algebra by using a deformation
parameter $h$ to get a relation between the Poisson bracket and the
commutator. Therefore we have the problem to find the deformation
of the Poisson algebra $(X(S,SL(2,\mathbb{C})),\left\{ \,,\,\right\} )$.
The solution to this problem can be found via two steps: 
\begin{enumerate}
\item at first find another description of the Poisson algebra by a structure
with one parameter at a special value and 
\item secondly vary this parameter to get the deformation. 
\end{enumerate}
Fortunately both problems were already solved (see \cite{Turaev1989,Turaev1991}).
The solution of the first problem is expressed in the theorem: 

The Skein module $K_{-1}(S\times[0,1])$ (i.e. $t=-1$) has the structure
of an algebra isomorphic to the Poisson algebra $(X(S,SL(2,\mathbb{C})),\left\{ \,,\,\right\} )$.\emph{
}(see also \cite{BulPrzy:1999,Bullock1999}) 

Then we have also the solution of the second problem: 

The skein algebra $K_{t}(S\times[0,1])$ is the quantization of the
Poisson algebra $(X(S,SL(2,\mathbb{C})),\left\{ \,,\,\right\} )$
with the deformation parameter $t=\exp(h/4)$.(see also \cite{BulPrzy:1999})\emph{ }

\begin{figure}
\begin{center}\includegraphics[scale=0.25]{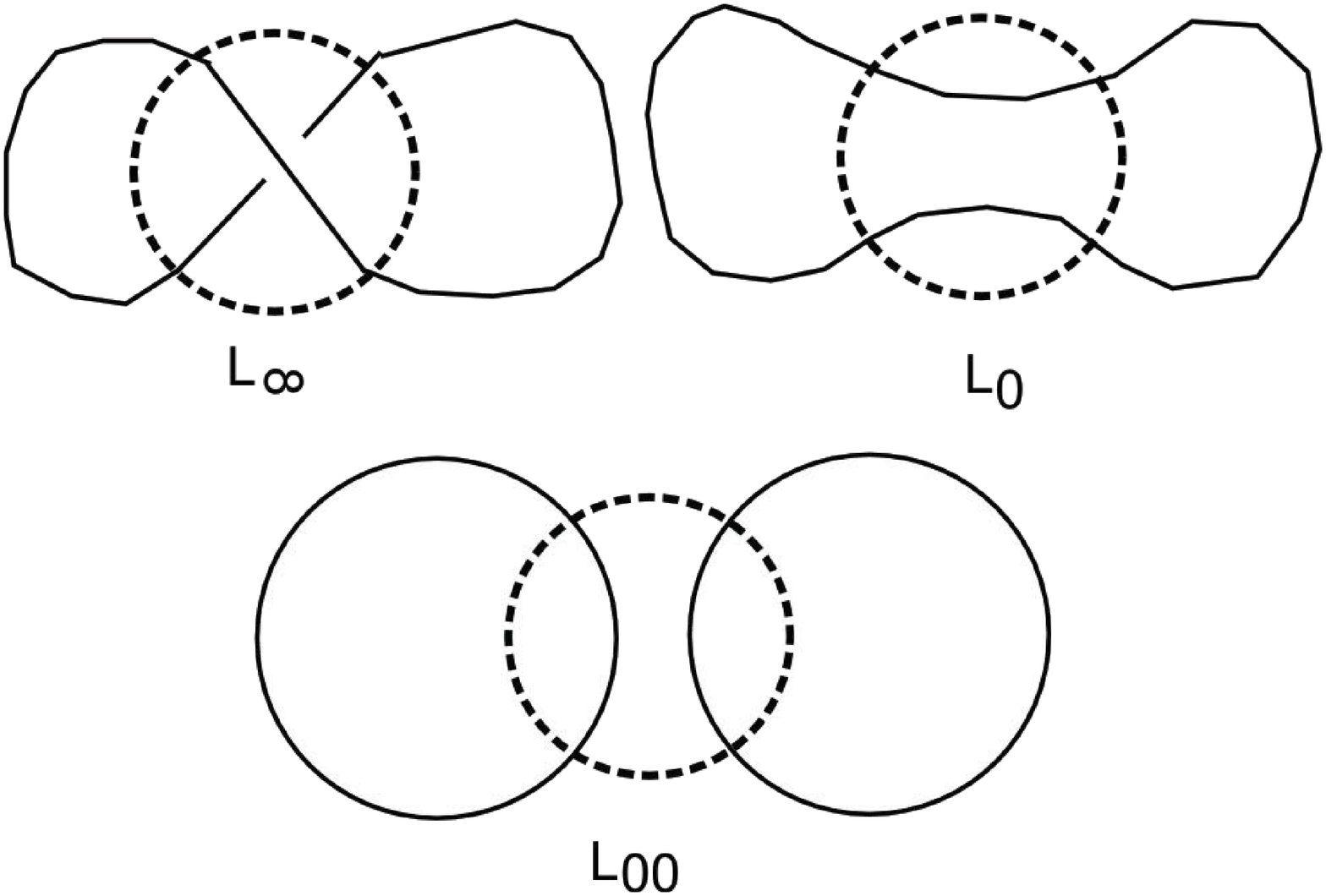}\end{center}

\caption{crossings $L_{\infty},L_{o},L_{oo}$\label{fig:skein-crossings}}
\end{figure}

To understand these solutions we have to introduce the skein module
$K_{t}(M)$ of a 3-manifold $M$ (see \cite{PrasSoss:97}). For that
purpose we consider the set of links $\mathcal{L}(M)$ in $M$ up
to isotopy and construct the vector space $\mathbb{C}\mathcal{L}(M)$
with basis $\mathcal{L}(M)$. Then one can define $\mathbb{C}\mathcal{L}[[t]]$
as ring of formal polynomials having coefficients in $\mathbb{C}\mathcal{L}(M)$.
Now we consider the link diagram of a link, i.e. the projection of
the link to the $\mathbb{R}^{2}$ having the crossings in mind. Choosing
a disk in $\mathbb{R}^{2}$ so that one crossing is inside this disk.
If the three links differ by the three crossings $L_{oo},L_{o},L_{\infty}$
(see figure \ref{fig:skein-crossings}) inside of the disk then these
links are skein related. 
Then in $\mathbb{C}\mathcal{L}[[t]]$ one writes the skein relation%
\footnote{The relation depends on the group $SL(2,\mathbb{C})$.%
} $L_{\infty}-tL_{o}-t^{-1}L_{oo}$. Furthermore let $L\sqcup O$ be
the disjoint union of the link with a circle then one writes the framing
relation $L\sqcup O+(t^{2}+t^{-2})L$. Let $S(M)$ be the smallest
submodul of $\mathbb{C}\mathcal{L}[[t]]$ containing both relations,
then we define the Kauffman bracket skein module by $K_{t}(M)=\mathbb{C}\mathcal{L}[[t]]/S(M)$.
We list the following general results about this module:
\begin{itemize}
\item The module $K_{-1}(M)$ for $t=-1$ is a commutative algebra.
\item Let $S$ be a surface then $K_{t}(S\times[0,1])$ caries the structure
of an algebra.
\end{itemize}
The algebra structure of $K_{t}(S\times[0,1])$ can be simple seen
by using the diffeomorphism between the sum $S\times[0,1]\cup_{S}S\times[0,1]$
along $S$ and $S\times[0,1]$. Then the product $ab$ of two elements
$a,b\in K_{t}(S\times[0,1])$ is a link in $S\times[0,1]\cup_{S}S\times[0,1]$
corresponding to a link in $S\times[0,1]$ via the diffeomorphism.
The algebra $K_{t}(S\times[0,1])$ is in general non-commutative for
$t\not=-1$. For the following we will omit the interval $[0,1]$
and denote the skein algebra by $K_{t}(S)$.

\subsection{Temperley-Lieb algebra and Alexanders horned ball}

Now we will present the relation between skein spaces and wild embeddings
(in particular to its $C^{*}-$algebra). For that purpose we will
concentrate on the wild embedding $i:D^{3}\to S^{3}$ or equivalently
$i:D^{2}\times[0,1]\to S^{3}$ of Alexanders horned ball. We will
explain now, that the complement $S^{3}\setminus i(D^{2}\times[0,1])$
and its fundamental group $\pi_{1}\left(S^{3}\setminus i(D^{2}\times[0,1])\right)$
can be described by closed curves around tubes (or annulus) $S^{1}\times[0,1]$. 

Let $C$ be the image $C=i(D^{2}\times[0,1])$ decomposed into components
$C_{i}$ so that $C=\cup_{i}C_{i}$. Furthermore, let $C_{i}$ be
the decomposition of $i(D^{2}\times[0,1]$) at $i$th level (i.e.
a union of $D^{2}\times[0,1]$). The complement $S^{3}\setminus C_{i}$
of $C_{i}$ with $n_{i}$ components (i.e. $C_{i}=\sqcup_{1}^{n_{i}}(D^{2}\times[0,1]$)
has the same (isomorphic) fundamental group like $\pi_{1}(\sqcup_{1}^{n_{i}}(S^{1}\times[0,1])$
of $n_{i}$ components of $S^{1}\times[0,1]$. Therefore, instead
of studying the complement we can directly consider the annulus $S^{1}\times[0,1]$
replacing every $D^{2}\times[0,1]$ component. 

Let $C'$ be the boundary of $C$, i.e. in every component we have
to replace every $D^{2}\times[0,1]$ by $S^{1}\times[0,1]$. The skein
space $K_{t}(S^{1}\times[0,1])$ is a polynomial algebra (see the
previous subsection) $\mathbb{C}[\alpha]$ in one generator $\alpha$
(a closed curve around the annulus). Let $TL_{n}$ be the Temperley-Lieb
algebra, i.e. a complex $*-$algebra generated by $\left\{ e_{1},\ldots,e_{n}\right\} $
with the relations
\begin{eqnarray}
e_{i}^{2}=\tau e_{i}\,, & e_{i}e_{j}=e_{j}e_{i}\,:\,|i-j|>1,\nonumber \\
e_{i}e_{i+1}e_{i}=e_{i}\,, & e_{i+1}e_{i}e_{i+1}=e_{i+1}\,,\, e_{i}^{*}=e_{i}\label{Jones-algebra}
\end{eqnarray}
and the real number $\tau$. If $\tau$ is the number $\tau=a_{0}^{2}+a_{0}^{-2}$
with $a_{0}$ a $4n$th root of unity ($a_{0}^{4k}\not=1$ for $k=1,\ldots,n-1$)
then there is an element $f^{(n)}$ with 
\begin{eqnarray*}
f^{(n)}A_{n} & = & A_{n}f^{(n)}=0\\
1_{n}-f^{(n)} & \in & A_{n}\\
f^{(n)}f^{(n)} & = & f^{(n)}
\end{eqnarray*}
in $A_{n}\subset TL_{n}$ (a subalgebra generated of $\left\{ e_{1},\ldots,e_{n}\right\} $
missing the identity $1_{n}$), called the Jones-Wenzl idempotent.
The closure of the element $f^{(n+1)}\in TL_{n+1}$ in $K_{t}(S^{1}\times[0,1]))$
is given by the image of the map $TL_{n+1}\to K_{t}(S^{1}\times[0,1]))$
which maps $f^{(n+1)}$ to some polynomial $S_{n+1}(\alpha)$ in the
generator $\alpha$ of $K_{t}(S^{1}\times[0,1])$). Therefore we obtain
a relation between the generator $\alpha$ and the element $f^{(n)}$
for some $n$. 

\begin{figure}
\includegraphics[scale=0.3]{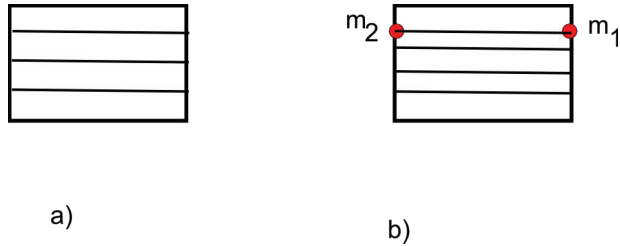}

\caption{Decomposition of $D^{2}$ (represented by rectangle) into smaller
rectangles (left fig.) and a representation for a closed curve by
a pair of opposite points (right fig.)\label{fig:Decomposition-D2}}

\end{figure}

Alexanders horned ball is homeomorphic to $D^{3}=D^{2}\times[0,1]$
(by definition of an embedding). The wilderness is given by a decomposition
of $D^{2}\times[0,1]$ into an infinite union of $(D^{2}\times[0,1])-$components
$C_{i}$ (in the notation above). By the calculation in subsection
\ref{sub:Example-Alexanders-horned-ball}, we have an infinite fundamental
group where every generator is represented by a curve around one $(D^{2}\times[0,1])-$components
$C_{i}$. This decomposition can be represented by a decomposition
of a square (as substitute for $D^{2}$) into (countable) infinite
rectangles (see Fig. \ref{fig:Decomposition-D2}a). 
 Every closed curve surrounding $C_{i}$ is a pair of opposite points
at the boundary (see Fig. \ref{fig:Decomposition-D2}b), the starting
point of the curve and one passing point (to identify the component).
Every $C_{i}$ gives one pair of points. Motivated by the discussion
above, we consider the skein algebra $K_{t}(D^{2},2n)$ with $2n$
marked points (representing $n$ components). This algebra is isomorphic
(see \cite{PrasSoss:97}) to the Temperley-Lieb algebra $TL_{n}$.
As Jones \cite{Jon:83} showed: the limit case $\lim_{n\to\infty}TL_{n}$
(considered as direct limit) is the factor $I\! I_{1}$. Thus we have
constructed the factor $I\! I_{1}$ algebra as skein algebra.

\emph{Therefore we have shown that the enveloping von Neumann algebra
}
\[
W(C,\pi_{1}(S^{3}\setminus A))
\]
\emph{ (=the hyperfinite factor $I\! I_{1}$ algebra) is obtained
by deformation quantization of a classical Poisson algebra (the tame
embedding). But then, a wild embedding can be seen as a quantum state.}

\section{Wild embeddings and 4-manifolds\label{sec:Wild-embeddings-exotic-4MF}}

This section is a kind of motivation that wild embeddings should be
considered. We start with the physically significant non-compact examples
of a spacetime: $S^{3}\times\mathbb{R}$. This non-compact 4-manifold
has the usual form used in general relativity (GR). There is a global
foliation along $\mathbb{R}$, i.e. $S^{3}\times\left\{ t\right\} $
with $t\in\mathbb{R}$ are the (spatial) leafs. $S^{3}\times\mathbb{R}$
with this foliation is called the ''standard $S^{3}\times\mathbb{R}$''.
But this choice is not unique. In dimension 4, there is a plethora
(uncountable many) of exotic smoothness structures (see \cite{Asselmeyer2007}).
In the following we will denote an exotic version by $S^{3}\times_{\theta}\mathbb{R}$.
The construction of $S^{3}\times_{\theta}\mathbb{R}$ is rather complicated
(see \cite{Fre:79}). As a main ingredient one needs a homology 3-sphere
$\Sigma$ (i.e. a compact, closed 3-manifold with the homology groups
of the 3-sphere) which does not bound a contractable 4-manifold (i.e.
a 4-manifold which can be contracted to a point by a smooth homotopy).
Interestingly, this homology 3-sphere $\Sigma$ is smoothly embedded
in $S^{3}\times_{\theta}\mathbb{R}$ (as cross section, i.e. $\Sigma\times\left\{ 0\right\} \subset S^{3}\times_{\theta}\mathbb{R}$).
What about the foliation of $S^{3}\times_{\theta}\mathbb{R}$? There
is no foliation along $\mathbb{R}$ but there is a codimension-one
foliation of the 3-sphere $S^{3}$ (see \cite{AsselmeyerKrol2009}
for the construction). So, $S^{3}\times_{\theta}\mathbb{R}$ is foliated
along $S^{3}$ and the leafs are $S_{i}\times\mathbb{R}$ with the
surfaces $\left\{ S_{i}\right\} _{i\in I}\subset S^{3}$. But what
happens with the 3-spheres in $S^{3}\times_{\theta}\mathbb{R}$? There
is no smoothly embedded $S^{3}$ in $S^{3}\times_{\theta}\mathbb{R}$
(otherwise it would have the standard smoothness structure). But there
is a wildly embedded $S^{3}$! This case is generic, i.e. also all
known exotic smoothness structures of non-compact 4-manifolds have
this property. Therefore, \emph{wild embeddings are generic for our
4-dimensional spacetime}.

\section{Conclusion}

In this paper we discussed a spacetime realization of a quantum state.
We conjectured that the quantum state is given by a wild embedding
like Alexanders horned ball. If this conjecture is true then we must
obtain a wild embedding by some quantization process from a classical
state. This program was done in this paper. At first we identify the
classical state with (the expected) tame embedding. Using Drinfeld-Turaev
quantization, we constructed a (deformation) quantum space and relate
them to a wild embedding. We demonstrate the process by the example
of Alexanders horned ball. Wild embeddings are generic constructions
in the theory of 4-manifolds with exotic smoothness structures. Therefore,
we conjecture that our technique must be useful in a future quantum
gravity theory.

At the end we will discuss a cosmological consequence of our approach.
Assume an exotic $S^{3}\times_{\theta}\mathbb{R}$ as a model of our
spacetime. A scaling of the $\mathbb{R}$ parameter leads to a beginning
at the $-\infty$-''point''. If we further assume a finite size of
the spatial component $S^{3}$ at the beginning (using a big bounce
effect, see \cite{AshPawSin:06}), then we have to look for 3-sphere
at the $-\infty$-''point''. But exotic smoothness enforce us: there
is no smoothly embedded 3-sphere and we obtain a wildly embedded 3-sphere
again. Therefore we obtain as a result of this paper: the cosmos started
in a quantum state (represented by a wildly embedded 3-sphere).

\end{document}